# Melting of $B_{12}P_2$ boron subphosphide under pressure

Vladimir L. Solozhenko,[1,*] Vladimir A. Mukhanov,[1] Petr S. Sokolov,[1] Yann Le Godec,[2] Kirill A. Cherednichenko,[1,2,3] Zuzana Konôpková[4]

[1] *LSPM–CNRS, Université Paris Nord, 99 avenue J.-B. Clément, 93430 Villetaneuse, France*

[2] *IMPMC-CNRS, UMR 7590, UPMC Sorbonne Universités, 4 place Jussieu, 75005 Paris, France*

[3] *Synchrotron Soleil, L'Orme des Merisiers, St. Aubin, 91192 Gif-sur-Yvette, France*

[4] *DESY Photon Science, Notkestrasse 85, 22607 Hamburg, Germany*

**Abstract**

Melting of boron subphosphide ($B_{12}P_2$) to 26 GPa has been studied by *in situ* synchrotron X-ray powder diffraction in a laser-heated diamond anvil cell, and by quenching and electrical resistance measurements in a toroid-type high-pressure apparatus. $B_{12}P_2$ melts congruently, and the melting curve has a positive slope of 23(6) K/GPa. No solid-state phase transition was observed up to the melting in the whole pressure range under study.

**Introduction**

Elemental boron and icosahedral boron-rich solids are important materials with remarkable properties that have a wide range of industrial applications [1]. Boron subphosphide $B_{12}P_2$ is a hard refractory compound with wide band gap and superior chemical resistance [2-4]. At near-ambient (10 MPa) pressure the melting temperature of $B_{12}P_2$ was evaluated as 2393 ± 30 K [4]. Melting behavior and phase stability of boron subphosphide at high pressure have not been studied so far. Here we report the pressure dependence of $B_{12}P_2$ melting temperature in the 2.6–26 GPa range from *in situ* and *ex situ* experiments.

---
[*] Corresponding author. E-mail: vladimir.solozhenko@univ-paris13.fr

**Experimental**

Polycrystalline single-phase stoichiometric boron subphosphide produced by self-propagating high-temperature reaction between boron phosphate and mixture of $MgB_2$ and metallic magnesium according to the method described elsewhere [5] was used in the melting experiments. The lattice parameters of as-synthesized $B_{12}P_2$ ($a$ = 5.982(20), $c$ = 12.00(2) Å) are in good agreement with literature data ($a$ = 5.9879, $c$ = 11.8479 Å [6]).

*In situ* experiments in the 16-26 GPa pressure range have been performed in a laser-heated diamond anvil cell (DAC) using angle-dispersive synchrotron X-ray diffraction at Extreme Conditions beamline P02.2 of the PETRA III storage ring (DESY) [7]. DACs with large optical aperture and 400-μm diameter culets were used. Small (40×40 μm, 10-15 μm thin) bulk samples of $B_{12}P_2$ were loaded in the cylindrical cavity of 100-μm diameter drilled in pre-indented to 50 μm thickness rhenium gasket, and the rest of the cavity was filled by KCl pressure medium insuring quasi-hydrostatic conditions at high temperature, with advantage of chemical inertness with regard to the B-P melt. The monochromatic X-ray beam (42 keV, $\lambda$ = 0.2898 Å) was focused down to 2 μm×4 μm. The diffraction patterns were recorded using XRD1621 (PerkinElmer) flat panel detector and integrated using Fit2D software [8]. Sample-detector distance (500 mm) was calibrated using $CeO_2$ NIST standard. Typical acquisition time was 30 seconds. The sample pressure during DAC compression has been determined using equation of state of KCl [9]. To estimate the pressure change upon heating, we used the approach described elsewhere [10]. The overall pressure uncertainty at high temperature was less than 2 GPa. Laser heating was performed in a double-sided off-axis infrared laser system (continuous fiber YAG laser, $\lambda$ = 1070 nm). Laser was focused down to 20 μm. Double-sided temperature measurements were accomplished through standard grey body radiation measurement via an Action spectrometer SP-2356 (Princeton Instruments). Temperature was measured before and after each acquisition of diffraction pattern, and the average values were used. Above 2000 K the temperature uncertainty can be estimated as ±50 K.

*In situ* experiments in the 1.6-5 GPa range have been performed using Paris-Edinburgh press and energy-dispersive X-ray diffraction at PSICHÉ beamline, synchrotron SOLEIL. Standard boron-epoxy gaskets with hBN pressure medium were used. Pressure and temperature were evaluated from equation of state of hBN at room temperature [11] and previously obtained temperature *vs* power calibration curve.

$B_{12}P_2$ melting in the 2.6–7.7 GPa pressure range has been studied by quenching using a toroid-type high-pressure apparatus with a specially designed high-temperature (up to 3500 K) cell. The details of the cell calibration against temperature and pressure are described earlier [12]. $B_{12}P_2$ powder was compacted into pellet and placed in a boron nitride (Saint-Gobain, grade AX05) capsule. No signs of chemical interaction between $B_{12}P_2$ and BN capsule were observed over the whole *p-T* range under study. After isothermal holding time of 60–90 s at the desired pressure and temperature, the sample was quenched by switching off the power and slowly decompressed down to ambient pressure. In some experiments no h-BN capsule has been used, and the appearance of liquid phase has been detected *in situ* by electrical resistance measurements using the method described elsewhere [13] (the sample was in a direct contact with graphite heater; no interaction between melt and graphite was observed).

The recovered samples have been studied by powder X-ray diffraction (Equinox 1000 Inel diffractometer; Cu Kα and Co Kα radiation), high-resolution scanning electron microcopy (Carl Zeiss Leo Supra 40VP and 50VP) and X-ray electron probe microanalysis (S440 Leica, Princeton Gamma-Tech and Supra 50VP, INCA Energy+ Oxford). Raman spectra were recorded using a Horiba Jobin Yvon HR800 spectrometer equipped with 632.8 nm He-Ne laser (10 μm beam).

**Result and discussion**

Fig. 1 shows a laser-heating sequence of X-ray diffraction patterns of boron subphosphide at 26 GPa. The melting was evident by complete disappearance of all $B_{12}P_2$ reflections at 3050 K. No diffuse X-ray scattering of the liquid was observed, probably due to an insufficient diffraction intensity from a very small sample. Reentering the solid phase from melt by decreasing temperature always resulted in reappearance of diffraction lines of $B_{12}P_2$.

The lattice parameters of the samples quenched from different pressures and temperatures are very close to the literature values, and lines of other phases (boron, phosphorus, BP, etc.) are absent in the diffraction patterns which is indicative of the congruent type of $B_{12}P_2$ melting under pressure. The results of SEM/EDX and Raman studies of the recovered samples also support this conclusion.

The grain texture of $B_{12}P_2$ samples quenched from different *p-T* conditions was investigated by scanning electron microscopy. The characteristic images of the samples recovered from temperatures below and above melting curve are shown in Fig. 2. The samples undergone full melting are dense, pore-free and exhibit extensive re-crystallization with grain growth up to several hundreds microns (Fig. 2b), while all non-melted samples are characterized by a grain size of a few microns (Fig. 2a).

Melting of boron subphosphide is accompanied by a drop in electrical resistance, and this resistance discontinuity was found to be reproducible upon several heating/cooling cycles. The specific electrical conductivity of the boron subphosphide melt at 2.6 GPa and 2500 K was evaluated as $13(6) \times 10^5\ \Omega^{-1} \cdot m^{-1}$, which virtually coincides with conductivity of silicon melt at ambient pressure ($13.2 \times 10^5\ \Omega^{-1} \cdot m^{-1}$ [14]), while the conductivity of solid $B_{12}P_2$ at this pressure ($0.32(16) \times 10^5\ \Omega^{-1} \cdot m^{-1}$ at 2400 K) is lower by 1.5 order of magnitude. At 6.1 GPa the difference in conductivity is less, i.e. $10(5) \times 10^5\ \Omega^{-1} \cdot m^{-1}$ (at 2620 K) and $1.2(6) \times 10^5\ \Omega^{-1} \cdot m^{-1}$ (at 2540 K) for melt and solid, respectively. Thus, upon solid–liquid phase transition boron subphosphide transforms from direct semiconductor (band gap of ~3 eV [1]) to metallic melt. Recently the similar phenomenon has been observed for boron and its carbide ($B_4C$), nitride (BN) and phosphide (BP). Under pressure the corresponding melts are good conductors with specific electrical conductivity values comparable with that of iron melt at ambient pressure [13].

The results obtained by *in situ* studies and quenching are in a good agreement (Fig. 3). $B_{12}P_2$ melting curve (dashed line in Fig. 3) has positive slope of 23(6) K/GPa that points to a lower density of the melt as compared to the solid phase at melting point. Taking into account that $B_{12}P_2$ is a typical boron-rich solid, we compared its melting behaviour with other compounds containing icosahedral $B_{12}$ structural units. Thus, at pressures to 8 GPa the melting slopes for β-rhombohedral boron (β-$B_{106}$), boron suboxide ($B_6O$) and boron carbide ($B_4C$) are 15.5 K/GPa [15], ~60 K/GPa [16,17] and -13 K/GPa [12], respectively. Such significant difference in melting slops of two borides ($B_6O$ and $B_4C$) with similar crystal structures derived from α-B is apparently due to the difference in structure and, hence, in density of the resulting melts. Another compound of the B–P binary system, cubic boron phosphide (BP) with zinc-blende structure, exhibits completely different melting behaviour under compression: the slope of the melting curve is significantly negative i.e. -60(7) K/GPa [18], most probably due to more efficient packing of atoms in the melt resulting from increase in coordination.

According to *in situ* X-ray diffraction data, at pressures to 26 GPa the icosahedral crystal structure of boron subphosphide remains stable up to the melting. Similar excellent phase stability is observed for other icosahedral boron-rich compounds, such as $B_6O$ [16,17] and $B_4C$ [12,19], while elemental boron exhibit several solid-state phase transitions in the same *p-T* range [15,20,21].

**Conclusions**

The melting curve of $B_{12}P_2$ boron subphosphide has been determined up to 26 GPa. The melting slope is positive, which is expected for boron-rich solids. Melting is of congruent nature, and the melt is metallic. The icosahedral crystal structure of $B_{12}P_2$ remained stable up to the melting in the whole pressure range under study (2.6–26 GPa).

**Acknowledgments**

The authors thank Dr. P. Munsch for assistance in DAC experiments, and Drs. O. Brinza and L.A. Trusov for SEM/EDX studies of recovered samples. *In situ* experiments at DESY have been carried out during beam time allocated to the Projects DESY-D-I-20090172 EC and DESY-D-I-20120021 EC and received funding from the European Community's Seventh Framework Programme (FP7/2007-2013) under grant agreement n° 226716. This work was financially supported by the Agence Nationale de la Recherche (grant ANR-2011-BS08-018).

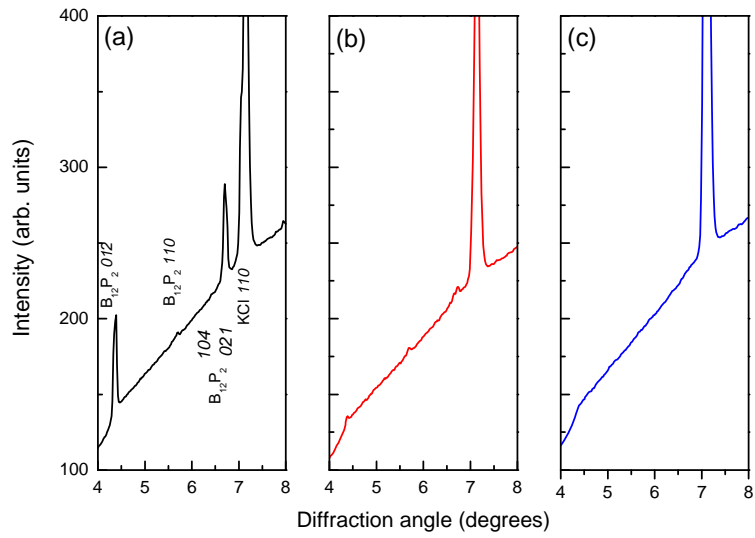

Fig. 1 Synchrotron X-ray diffraction patterns (λ = 0.2898 Å) of $B_{12}P_2$ in KCl pressure medium at 26 GPa and different temperatures. Black pattern (a) is taken at 2930 K, just below the $B_{12}P_2$ melting point; red pattern (b) corresponds to the partial melting at 3020 K; and blue pattern (c) is taken at 3050 K, just above the melting point (only lines of the solid pressure medium can be seen).

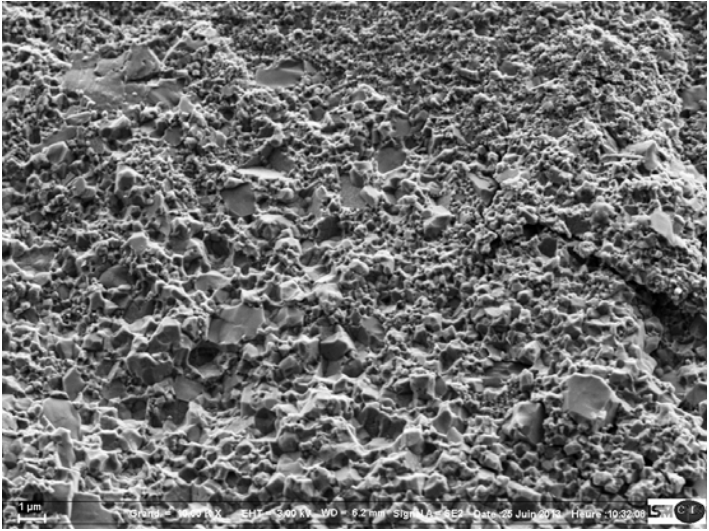     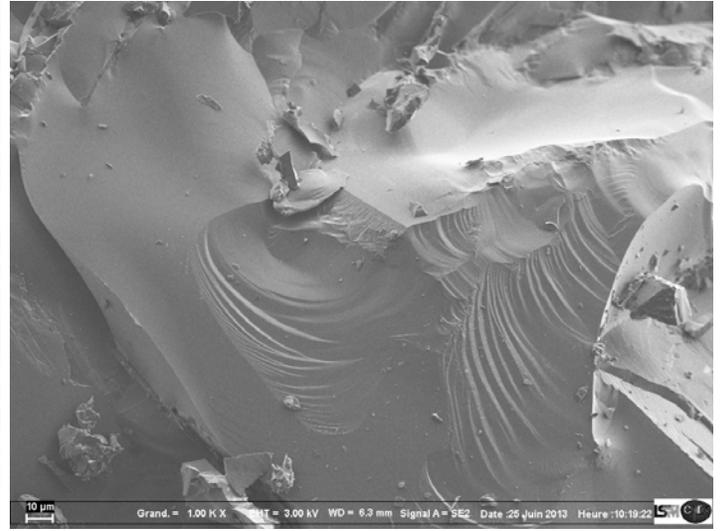

(a)                                           (b)

Fig. 2   Scanning electron microscopy images of $B_{12}P_2$ samples quenched from (a) 5.2 GPa/2500 K (10k×), not melted; and (b) 7.5 GPa/2723 K (1000×), melted.

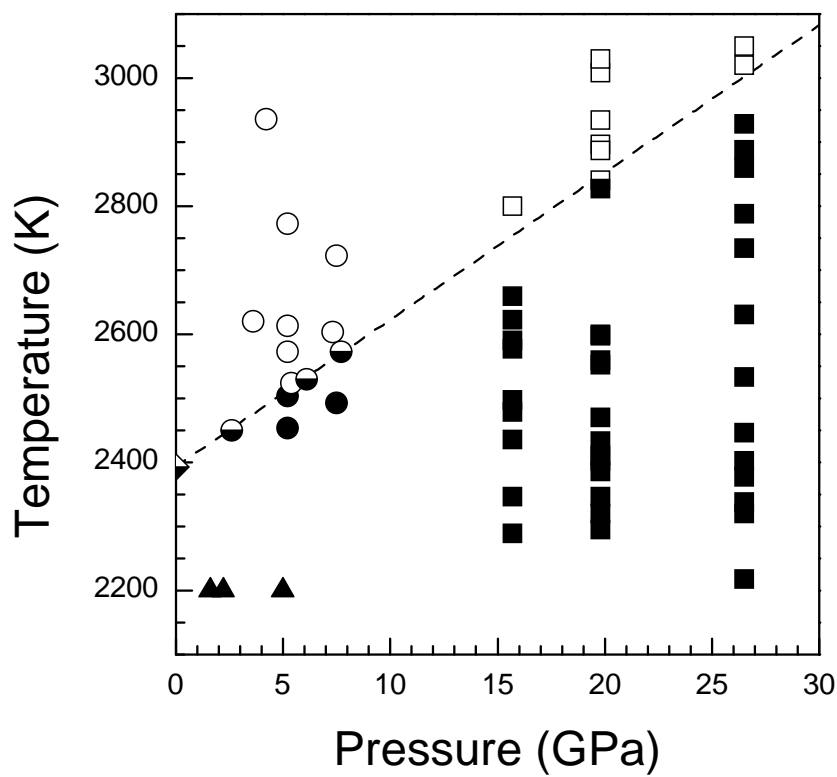

Fig. 3  Melting of $B_{12}P_2$ under pressure. The half-filled diamond indicates the melting temperature at 10 MPa (2393±30 K [4]); circles show the results of quenching experiments; squares show the results of *in situ* experiments in DACs; triangles present *in-situ* data obtained in Paris-Edinburgh press. The open symbols correspond to the melting, solid ones – to its absence; half-filled circles are the melting onset registered *in situ* by electrical resistance measurements. Temperature error bars (not shown) are ±20 K for the large-volume data points and ±50 K for the DAC data points. Dashed line is the linear approximation of the melting curve.